\pdfoutput=1
\pdfoutput=1
\pdfoutput=1
\pdfoutput=1
\pdfoutput=1

\documentclass[fleqn]
{article}

\usepackage[T1]{fontenc}
\usepackage{amsfonts,amssymb,amsmath} % for math symbols.
\usepackage{bm}% bold math
\usepackage[all]{xy}
\usepackage{graphics,graphicx,epsfig}
\usepackage{color}
\usepackage{fancyhdr}
\fancypagestyle{plain}{%
 \fancyhead{} % ----на обычных страницах без колонтитулов

}
\usepackage [ english
] { babel }
\usepackage{makeidx}
\makeindex

\usepackage{xr-hyper}
\usepackage[unicode,pagebackref]{hyperref}
\hypersetup{pdftex}

\newcommand{\beq}{\begin{equation}}
\newcommand{\eeq}{\end{equation}}
\newcommand{\beqa}{\begin{eqnarray}}
\newcommand{\eeqa}{\end{eqnarray}}

 \title{ Embedding of binary image in the Gray planes
 }
 \author{Valery N. Gorbachev, Lev A. Denisov\footnote{E-mail: denisovlev@yandex.ru}, Elena M. Kainarova}

\date{\emph{\small {North - Western Institute of Printing}}\\
\emph{\small{St.-Petersburg State University of Technology and Design}} }

\begin{document}
\maketitle
\begin{abstract}
For watermarking of the digital grayscale image its Gray planes have been used.
With the help of the introduced representation over Gray planes the LSB embedding method and detection have been discussed.
It found that data, a binary image, hidden in the Gray planes
is more robust to JPEG lossy compression than in the bit planes.
\end{abstract}

%\tableofcontents
\section{Introduction}

The Gray code is well-known in the theory of information \cite{r1}. It allows to reveal one of the main features of digital images, its redundancy that is of interest of the modern steganography.

 The Gray planes of digital image obtained from its bit planes represent image in spatial domain. The great variety of the spatial domain watermarking techniques (see, for example, \cite{r2}) include LSB (Least Significant Bit) method, block and additive embedding
 for which binary representation is used.

Instead of bits there might be used trits achieved by pseudo-ternary encoding, that was developed for watermark \cite{r4}. It is possible hiding information in audio
files and it requires its own technique \cite{r4.2}.

After the data is embedded, the achieved stego work frequently would be storied in some graphic format for further usage. It can destroy the watermark because some graphical formats as JPEG use lossy compression.

 This problem is well-known. A simple solution is storing in a lossless format like TIFF or PNG. However JPEG is widely used in practice, so the robust to JPEG-compression steganography is much attended. Robust to lossy compression technique allows to achieve different problems including pattern recognition, image enhancement \cite{r5}
  and many others. Watermarking of JPEG file and its subsequent JPEG-compression is known as J2J (JPEG-to-JPEG) transform \cite{r6}. There is not unique solution for J2J and a large number of methods includes a trade-offs between hight compression level and watermark degradation.
\\
\\
In our work embedding of binary image into the Gray planes of a grayscale image under J2J transform considered.
 The Gray plane watermarking has been proposed in ref. \cite{r7} and developed by many authors \cite{r8, r8.1, r8.2}, it showed its resistance to RS analysis \cite{r9}, the $\chi^{2}$ attack \cite{r10} and SPA (Sample Pair Analysis) \cite{r11, r11.1}. For steganoanalysis the SPA watermark detectors may be very efficiency \cite{r12}, however its design is very complicated if the bit planes beginning from the second one are used \cite{r13}. In our work the fourth Gray plane is watermarked, for this case detector is unknown to authors. In our paper we focus on the Gray plane watermarking that is robustness to JPEG compression.

 The paper is organized as follows. First, hiding a binary image in the bit and Gray planes is considered and various detection procedures are introduced. Then a scheme included the lossy compression is analyzed and the distortion measures as $PSNR$ and others are calculated.

\section{Embedding and detection algorithms}

The Gray code for pixel \(c\) is created from its binary representation \(c=b_W,b_{W-1},\ldots,b_1\) as follows \(g_W=b_W\), \(g_V=b_V+b_{V+1},\ V<W\), where all \(b_V,g_V\in 0,1\). From here and later all bits are added by modulo two. Binary digit \(b_V\) has its weight \(2^{V-1}\), \(V=1,\ldots,W\), where the most significant bit is $b_{W}$ and LSB is $b_{1}$. It is not true for the Gray code that has the unnatural weights.

 The set \(B_V\) from digits \(b_V\) of all pixels is named bit plane, that is a binary image. As well the set $G_{V}$ of the codes \(g_V\) is named Gray plane, which is also a binary image. Any grayscale image can be represented over the bit and Gray planes. Let $C$ be a 8-bit grayscale image, then its representation over the bit planes has the forme
\(C=\sum_{V=1}^{8}{2^{V-1}B_V}\).
From definition of the Gray codes it follows that
\begin{eqnarray}
	\label{gray2bit}
	&B_8=G_8, \nonumber \\
	&B_T=G_8+\ldots+G_V+\ldots+G_T,
\end{eqnarray}
where \(T<8\).
Two important points may be made from these Eqs.
First, any modification of a Gray plane, say \(G_5\), results in modification of the bit plane set $\{B_1, \ldots, B_5\}$. Second,
representation over the Gray planes has the form
\cite{14}
\begin{equation}
	\label{gray2bit_full}
	C = 2^{7}G_{8}+\sum_{V=1}^{7}2^{V-1}(G_8+\ldots+G_V).
\end{equation}
 Consider the problem of hiding a binary image $M$ into cover work image $C$
 by watermarking its Gray planes. Assume the embedding algorithm be a bit addition
\[
	\label{method_not_blind}
	G_V \rightarrow G_{VM} = G_V + M,
\]
Indeed, the least significant planes of any image look as a noise being a random binary arrays so that they can be considered as a secrete key of the embedding algorithm (\ref{method_not_blind}).
As a result stego work reads
\begin{equation}
	\label{eq_bit}
	S = 2^7B_8+\ldots+2^{V-1}B_{VM}+\ldots+2^0B_{1M},
\end{equation}
where \(B_{TM}, T=1,\ldots,V\) are bit planes with hidden image \(M\): \(B_{VM}=B_V+M\).
 This Equation tells, that embedding into a Gray plane is equivalent to embedding into a set of bit planes $\{B_1,\ldots,B_V\}$. It is possible two ways of detection,
 the image $M$ can be extract from both the bit plane and the Gray plane.

Lets consider a non-blind detection for which the cover image \(C\) is required. The hidden image can be obtained from the Gray plane \(G_V\) by addition \(M = G_{VM} + G_V\). Introduce a function \(bitget(A,V)\), it is known in Matlab
and it calculates a bit plane \(V\) of an image \(A\). Then detection procedure using the Gray plane takes the forme
\begin{eqnarray}
	\label{detection_1}
	D_1: M = bitget(C,V)+bitget(C,V+1)+ %\nonumber\\
	bitget(S,V)+bitget(S,V+1).
\end{eqnarray}
where the bit addition results in
$
	bitget(C,V+1)+bitget(S,V+1)=0,
$
because the plane \(V+1\) is not watermarked.
In the same time image $M$ can be extracted from the bit planes
\(M=B_T+B_{TM}\), and next detection procedure is possible
\begin{equation}
	\label{detection_2}
	D_2: M = bitget(C,T) + bitget(S,T),
\end{equation}
where \(T=1,2,\ldots,V\).

The image redundancy allows to modify several bit and Gray planes without introducing noticeable changing. This fact leads to the blind detection for which the cover image is not required. Next scheme can implement this solution. The plane \(V\) is replaced with the plane \(K\) including \(M\) and to get $M$
the stego work planes \(K\) and \(V\) are extracted.
For the Gray planes this scheme has the next steeps.
\begin{itemize}
 \item Embedding
\[
	\label{method_blind}
	G_V \rightarrow G_{VM} = G_K + M.
\]
 \item Detection
\begin{eqnarray}
	\label{blind_detection_scheme}
	M = G_{VM}+G_K = bitget(S,V) + bitget(S,V+1)+ %\nonumber\\
	 bitget(S,K)
+bitget(S,K+1).
\end{eqnarray}
\end{itemize}
For particular case that is interesting for practice \(K = V+1\), we have
$
	M=bitget(S,V)+bitget(S,V+2).
$

In contrast to non blind detection the presented scheme can't be rewritten
with the use of the bit plane embedding options (\ref{eq_bit}). So
two detection procedures (\ref{detection_1}) and (\ref{detection_2}) are impossible. It can be illustrated by the next example. Assume \(V=2, K=3\) and \(G_2 \rightarrow G_{2M}=G_3+M\). Rewrite (\ref{gray2bit_full}) in the form
\[C=Z+2^2(Y+G_3)+2^1(Y+G_3+G_2)+2^0(Y+G_3+G_2+G_1),\]
where \(Y=G_8+\ldots+G_4\). Using expression for \(G_{2M}\), then instead of (\ref{gray2bit_full}) we find
\[S=Z+2^2(Y+G_3)+2^1(Y+M)+2^0(Y+G_1+M).\]
According to (\ref{blind_detection_scheme}) the detection is defined by expression \(M=bitget(S,2)+bitget(S,4)\).

\section{Lossy compression}
The JPEG format includes lossy compression, this is an complicated procedure we will describe by mapping
\[
	A\rightarrow A_q,
\]
where \(q=1,2,\ldots,100\) is named a quality parameter. It refers to the JPEG quantization matrixes. The lager \(q\) is the higher the compressed image quality is.

Next scheme includes lossy compression is considered.
 A binary image \(M\) is embedded into a bit plane \(B_V\) and a Gray plane \(G_V\) of a grayscale cover \(C\). Two obtained stego works \(S_B\) and \(S_G\) also cover image are compressed
\[
	M,C\rightarrow S_B,S_G,
\]
\[
	C,S_B,S_G\rightarrow C_q,S_{Bq},S_{Gq}.
\]
To extract $M$ it is possible four detection procedures namely
the blind and non-blind detection, also \(D_1\) and \(D_2\) procedures. Then we have four watermarks: \(M_b, M_{gb}, M_g, M_{gc}\) defined as
\[M_b=bitget(C_q, V)+bitget(S_{Bq}, V) \]
\[M_{gb}=bitget(C_q, T)+bitget(S_{Gq}, T) \]
\[M_g=bitget(C_q, V)+bitget(S_{Gq}, V) \]
\[M_{gc}=bitget(S, V)+bitget(S, V+2),\]
where \(T=1,\ldots,V\). Here watermark \(M_b\) is obtained after embedding and detection from a bit plane by non-blind detection, \(M_{gb}\) and \( M_g\) are obtained after embedding into a Gray plane by detection from a bit plane and
from a Gray plane by non-blind detection, finally \(M_{gc}\) is obtained after embedding into a Gray plane by blind detection.

\section{Experiment}

The main goal of our experiment is to compare watermarking of the bit planes and Gray planes and to find efficient detection.

For this purpose the distortion measures \(d(M,X)\) between the initial watermark \(M\) and the extracted after the compression watermarks \(X=M_b,M_{gb},M_g,M_{gc}\) have calculated. It has been used the Euclidean distance \(e(M,X)\), Peak Signal-Noise Ratio \(PSNR(M,X)\) and relative entropy \(Q(M||X)\). For two matrices \(M\) and \(X\) these measures are defined as follows \begin{eqnarray}
	\label{measures}
	 &e(M,X)=\sqrt{ (1/\Omega)\sum_{m,n}(M_{mn}-X_{mn})^2},\nonumber\\
	&PSNR(M,X)=20\log_{10}\Big(\max |M_{mn}|/e(M,X)\Big), \\
	 &Q(M||X)=\sum_{j}{p_M[j](\log_{2}{p_M[j]}-\log_{2}p_X[j])},\nonumber
\end{eqnarray}
where \(\Omega\) is a total number of elements, \(p_M[j],p_X[j]\) are histograms of brightness of \(M\) and \(X\).

\subsection{Embedding in the bit and the Gray planes}

\begin{figure}
  \includegraphics[trim=5cm 0 0 0, clip, width=16cm]{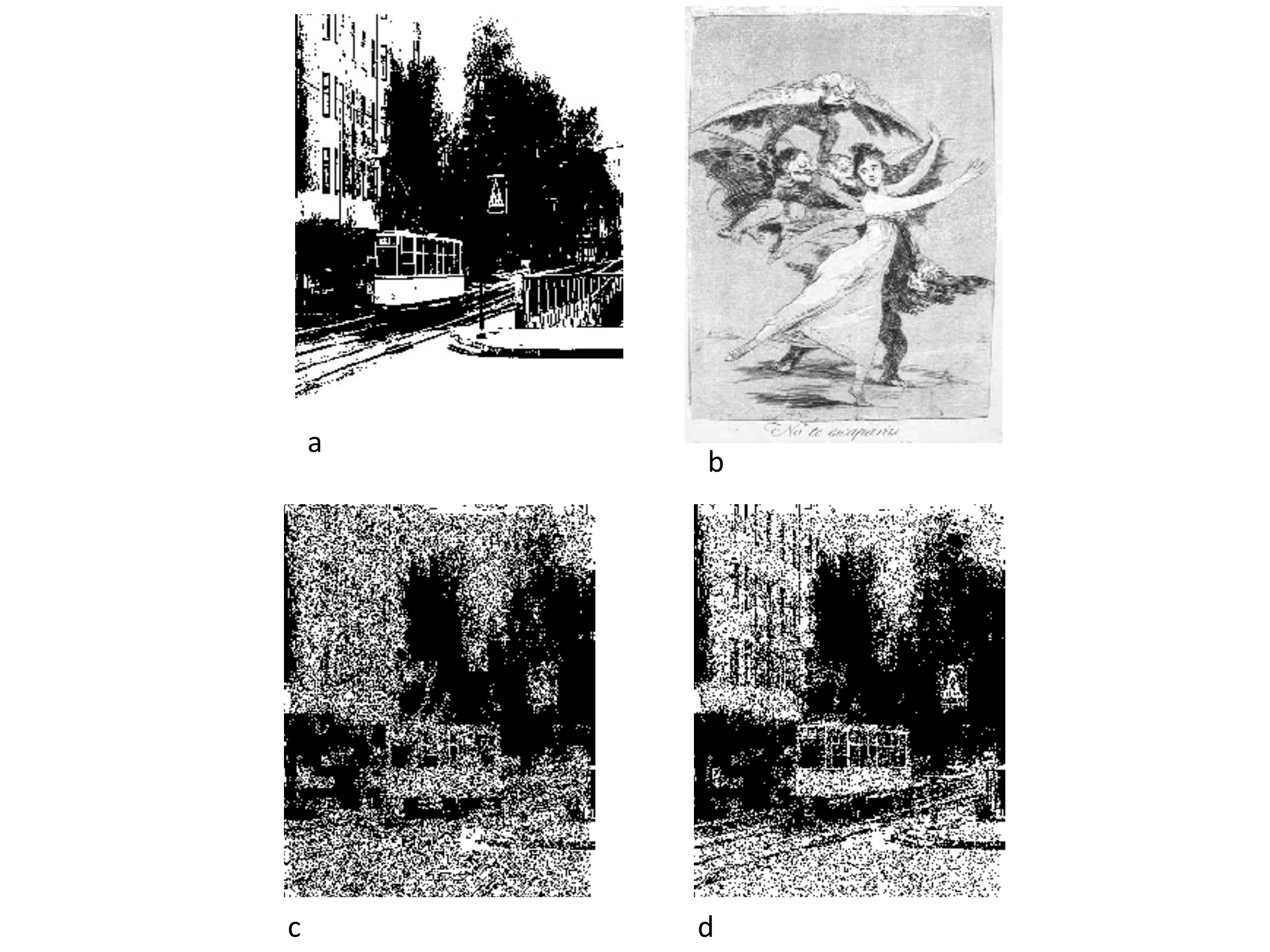}\\
  \caption{
  Embedding into the bit plane and the Gray plane: the binary watermark (a); the grayscale image (b); the watermark detected from the bit plane, \(q=90\) (c); the watermark detected from the Gray plane, \(q=90\) (d).
  }
  \label{figu1}
\hyperlink{figu1}{W}
\end{figure}

The binary image \(M\) and the grayscale cover work \(C\) are represented
 \hypertarget{figu1}
on the Fig.~\ref{figu1}. The fourth bit plane \(B_4\) and the fourth Gray plane \(G_4\) were selected. This choice results from the next trade off.

\begin{figure}
  \includegraphics[%height=14cm,
  width=16cm]{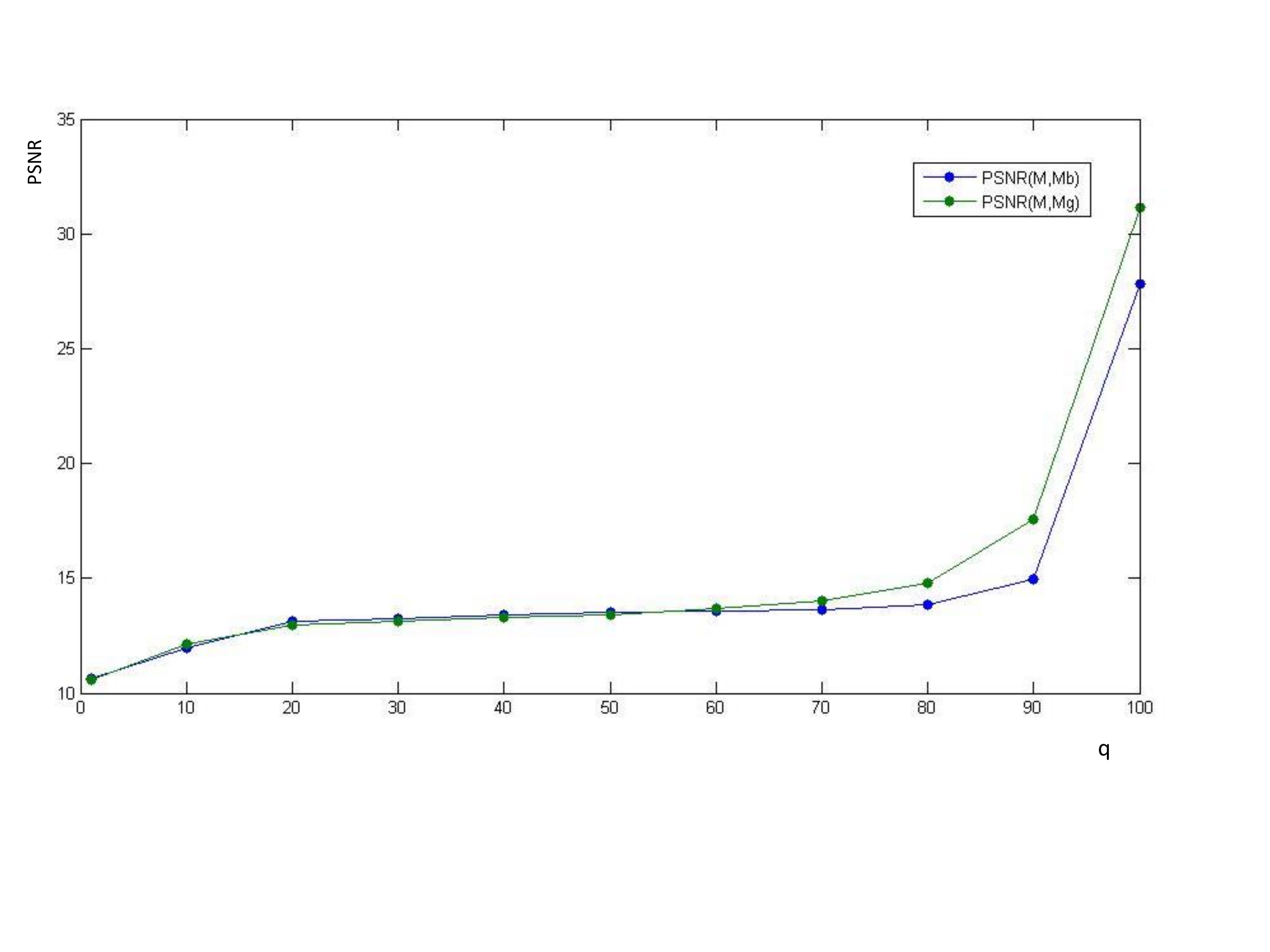}\\
  \caption{Peak Signal-Noise Ratio for embedding into the bit plane and the Gray plane.}
    \label{figu2}
\hyperlink{figu2}{W}
\end{figure}

The less significant bit plane is the higher distortion level due from lossy compression is and it would be better to hide data into more significant planes
in which the hidden data becomes visible. In our case if \(V=4\) the cover and the stego image are visually indistinguishable. Note that for \(V=4\) the simple attacks like \(\chi^2\) might be ineffective.
Fig.~\ref{figu2} shows watermarks \(M_b\) and \(M_g\) extracted from
the bit plane and the Gray plane after compression with the quality parameter \(q=90\). These watermarks are obtained by non-blind detection. The
 \(M_g\) has better visual quality and better distortion measures. For example, Fig.~\ref{figu2} presents PSNR averaged on 80 images. Indeed, these images are from the collection of Caprichos by de Goya. They have a complicated texture and it allows watermarking of significant bit planes. For the quality parameter \(q>50\), this range has the most interest, PSNR of \(M_g\) is in the range from 15 to 30 \(dB\) that refers to satisfactory visual quality.

The found results shows that Gray planes are more preferable for watermarking if the image would be store in jpeg file.

\subsection{Detection from the bit plane and the Gray plane}

\begin{figure}
  \includegraphics[%height=14cm,
  width=16cm]{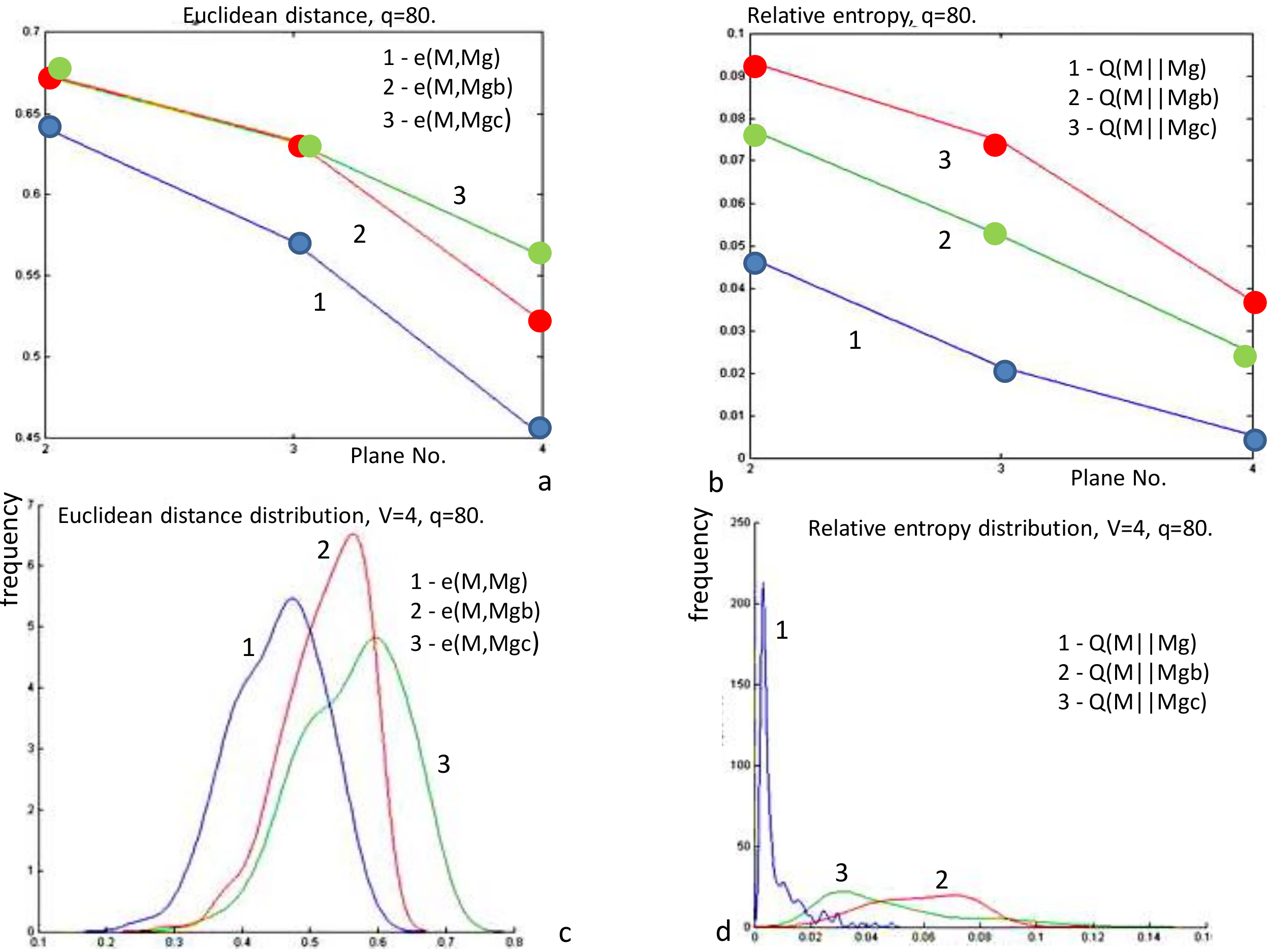}\\
  \caption{
  Embedding into the bit plane and the Gray plane: the binary watermark (a); the grayscale image (b); the watermark detected from the bit plane, \(q=90\) (c); the watermark detected from the Gray plane, \(q=90\) (d).
  }
  \label{figu3}
\hyperlink{figu3}{W}
\end{figure}

Here we consider the blind and non-blind detection
 for set of 200 grayscale images. The watermark was embedded in the Gray planes planes \(G_2, G_3, G_4 \) with \(q=70,80,90\) and the set of distortion measures \(d(M,X)\) defined by (\ref{measures}) were computed. Fig.~\ref{figu3} shows Euclidean distance and relative entropy averaged on 200 images for various \(V\). All functions are decreasing monotonically. The treason it that
 the less significant plane is the larger its redundancy destroyed by the compression.
 The measures are represented for the quality parameter \(q=80\), which is not the most preserving mode of compressing. It follows that the
 the non-blind detection has better distortion measures
 between \(M\) and \(M_{g}\) as for Euclidean distance and relative entropy.
 This is true for \( PSNR\) also.
 \hypertarget{figu3}
Fig.~\ref{figu3} presents two histograms of Euclidean distance and relative entropy. They describe distribution of its value found from the base of 200 images. These distributions are well distinguishable distributions, it means that the embedding methods have significant difference.

It follows from the found results that non-blind detection shows less distortion
of the extracted watermarks after compression.

\section*{Conclusion}
The Gray planes of digital image can be suitable for hiding information. In spatial domain the LSB embedding has the same features as for bit as for Gray planes
and results in difference if the image will be storied in a graphical format that has lossy compression as JPEG. There are various detection procedures to extract the hidden data after compression, particular a blind detection is possible. By analyzing a set of distortion measures as Euclidian distance, \(PSNR\), relative entropy it has been found that embedding and detection of watermark from the Gray planes is better than for the case of bit planes.

\newpage
%\input{Figu}
%%%%%%%%%%%%%%%%%%
%%%%%%%%%%%%%%%%%%%
%%%%%%%%%%%%%%

\end{document}